\documentclass[12pt]{article}

\usepackage{amsmath}
\usepackage{amssymb}
\usepackage{epsfig}

\newcommand{\capdef}{}
\newcommand{\mycaption}[2][\capdef]{\renewcommand{\capdef}{#2}%
        \caption[#1]{{\itshape #2}}} 
\makeatletter
\renewcommand{\fnum@table}{\textbf{\tablename~\thetable}}
\renewcommand{\fnum@figure}{\textbf{\figurename~\thefigure}}
\makeatother

\def\ltap{\ \raisebox{-.4ex}{\rlap{$\sim$}} \raisebox{.4ex}{$<$}\ }

\newcounter{myenumi}

\renewcommand{\themyenumi}{\roman{myenumi}}
{\end{list}}

\newlength{\myem}
\newcommand{\sep}[1]{#1}
\newcounter{mysubequation}[equation]
\renewcommand{\themysubequation}{\alph{mysubequation}}
\newcommand{\mytag}{\stepcounter{mysubequation}%
\tag{\theequation\protect\sep{\themysubequation}}}
\newcommand{\globallabel}[1]{\refstepcounter{equation}\label{#1}}

\makeatletter
\renewcommand{\section}{\@startsection{section}{1}{0em}{-\baselineskip}%
{\baselineskip}{\normalfont\large\bfseries}}
\renewcommand{\subsection}%
{\@startsection{subsection}{2}{0em}{-0.7\baselineskip}%
{0.7\baselineskip}{\normalfont\bfseries}}
\makeatother

\newcommand{\bea}{\begin{eqnarray*}}
\newcommand{\eea}{\end{eqnarray*}}

\newcommand{\GeV}{\,\mathrm{GeV}}

\newcommand{\eV}{\,\mathrm{eV}}

\newcommand{\reu}{{\nu_e\rightarrow\nu_\mu}}
\newcommand{\reub}{{\bar{\nu}_e\rightarrow\bar{\nu}_\mu}}
\newcommand{\ruu}{{\nu_\mu\rightarrow\nu_\mu}}
\newcommand{\ruub}{{\bar{\nu}_\mu\rightarrow\bar{\nu}_\mu}}
\newcommand{\rut}{{\nu_\mu\rightarrow\nu_\tau}}

\newcommand{\ret}{{\nu_e\rightarrow\nu_\tau}}

\newcommand{\NKT}{{N_{\text{kT}}}}
\newcommand{\dm}[1]{{\Delta m^2_{#1}}}

\def\ltap{\ \raisebox{-.4ex}{\rlap{$\sim$}} \raisebox{.4ex}{$<$}\ }

\begin{document}


\begin{titlepage}

\renewcommand{\thefootnote}{\alph{footnote}}

\ \vspace*{-3.cm}
\begin{flushright}
  {\hfill TUM--HEP--373/00}\\
  {\ }
\end{flushright}

\vspace*{0.5cm}

\renewcommand{\thefootnote}{\fnsymbol{footnote}}
\setcounter{footnote}{-1}

{\begin{center} {\Large\bf Extracting Matter Effects, Masses 
        and Mixings \linebreak at a Neutrino 
        Factory$^*$\footnote{\hspace*{-1.6mm}$^*$Work supported by 
        "Sonderforschungsbereich 375 f\"ur Astro-Teilchenphysik" der 
        Deutschen Forschungsgemeinschaft.}}
\end{center}}
\renewcommand{\thefootnote}{\alph{footnote}}

\vspace*{.8cm}
{\begin{center} {\large{\sc 
                M.~Freund\footnote[1]{\makebox[1.cm]{Email:}
                Martin.Freund@physik.tu-muenchen.de},~  
                P.~Huber\footnote[2]{\makebox[1.cm]{Email:}
                Patrick.Huber@physik.tu--muenchen.de}~and~
                M.~Lindner\footnote[3]{\makebox[1.cm]{Email:}
                Manfred.Lindner@physik.tu--muenchen.de}~
                }}
\end{center}}
\vspace*{0cm}
{\it 
\begin{center}  
                Theoretische Physik, Physik Department, 
                Technische Universit\"at M\"unchen,\\
                James--Franck--Strasse, D--85748 Garching, Germany
\end{center} }

\vspace*{1.5cm}

{\Large \bf 
\begin{center} Abstract \end{center}  }
We discuss and quantify different possibilities to determine 
matter effects, the value and the sign of $\dm{31}$, as well
as the magnitude of $\sin^2 2\theta_{13}$ in very long 
baseline neutrino oscillation experiments. We study neutrino
oscillation at a neutrino factory in the $\nu_\mu\rightarrow\nu_\mu$ 
disappearance and $\nu_e\rightarrow\nu_\mu$ appearance channels 
with and without muon charge identification. One possibility is 
to analyze the $\nu_e\rightarrow\nu_\mu$ appearance channels 
leading to wrong sign muon events, which requires however very 
good muon charge identification. Without charge identification 
it is still possible to operate the neutrino factory both with 
$\mu^-$ and $\mu^+$ beams and to analyze the differences in 
the total neutrino event rate spectra. We show that this leads 
already to a quite good sensitivity, which may be important if 
right sign charge rejection capabilities are insufficient.
With muon charge identification one can study the 
$\nu_\mu\rightarrow\nu_\mu$ disappearance and the 
$\nu_e\rightarrow\nu_\mu$ appearance channels independently. 
The best method is finally achieved by combining all available 
information of the $\nu_\mu\rightarrow\nu_\mu$ disappearance and 
$\nu_e\rightarrow\nu_\mu$ appearance channels with charge 
identification and we show the sensitivity which can be achieved. 

\vspace*{.5cm}

\end{titlepage}

\newpage

\renewcommand{\thefootnote}{\arabic{footnote}}
\setcounter{footnote}{0}


\section{Introduction \label{sec:SEC-intro}}

Proposed neutrino factories 
\cite{GEER98,gavela-c,BARGER99,BCR9905240,lyon} 
offer unique possibilities to improve the knowledge about neutrino 
masses, leptonic mixings and CP violation. The leptonic sector is 
not plagued by hadronic uncertainties such that vacuum neutrino 
oscillation allows to determine many parameters quite precisely.
Vacuum neutrino oscillation is however only sensitive to mass squared 
differences thus the sign of $\dm{31}$ can not be determined allowing 
therefore at the moment different scenarios for the ordering of mass 
eigenvalues. It has however been pointed out recently 
\cite{BARGER99b,BARGER2000,gavela-d} that the sign of $\dm{31}$ can 
be determined from the $\reu$ and $\reub$ appearance rates via matter 
effects \cite{MSW} in very long baseline neutrino oscillation 
experiments at neutrino factories. We discuss and quantify in this 
paper further possibilities to determine the sign of $\dm{31}$ 
via matter effects and $\sin^2 2\theta_{13}$ by considering
the appearance channels and/or the $\ruu$ and $\ruub$ 
disappearance channels in different scenarios with and 
without muon charge identification.
Altogether there are four possibilities. First, without charge 
identification one can not study wrong sign muon events (i.e.
the appearance channels), but one can operate a neutrino factory 
already both with $\mu^-$ and $\mu^+$ beams and analyze the 
differences in the combined muon neutrino and antineutrino 
event rate spectrum. This is useful if charge identification 
is not available or not operative in an initial stage of the 
experiment. We show that this leads already to a quite good 
sensitivity since the effects in the combined appearance and 
disappearance channels are for a baseline of $7332$~km about the same 
size and go in the same direction. 
Next, with muon charge identification, one can study 
the $\nu_\mu\rightarrow\nu_\mu$ disappearance and the 
$\nu_e\rightarrow\nu_\mu$ appearance channels separately.
For a baseline between $2800$~km and $7332$~km we find that the 
significance of these result is essentially unchanged compared to
7332~km. The point is that the total event rates 
decrease slower than the  vacuum rates leading for a large 
range to a statistically constant result. We use therefore
in this paper mostly a baseline of $7332$~km since for this 
larger baseline there is further information in the disappearance
channels. A third option would be to use the disappearance channel
alone, this provides however not much extra information. Last, the best 
method is achieved with muon charge identification by combining all 
available information of the $\nu_\mu\rightarrow\nu_\mu$ disappearance 
and $\nu_e\rightarrow\nu_\mu$ appearance channels. We show for our 
baseline of $7332$~km the sensitivity which can be achieved. 

The basic effect which allows to extract the sign of $\dm{31}$
comes from coherent forward scattering of electron neutrinos 
in matter which leads to effective masses and mixings different 
from vacuum. We will discuss in this paper a full three neutrino 
framework in the limit where the small solar mass splitting can 
be neglected in vacuum, i.e. $\dm{21}=0$. This degeneracy is
however destroyed in matter and we need therefore a full three neutrino 
description. The basic mechanism which allows the extraction 
of the sign of $\dm{}$ via matter effects can, however, already be 
seen in a simplified 2x2 picture with two mass eigenvalues $m_i$, 
$m_j$, $\dm{}=m_j^2-m_i^2$ and one mixing angle $\theta$ only. 
\newpage
One obtains then the well known relations for the parameters
in matter
\begin{equation}
\dm{m} = \dm{}C_\pm = \dm{}
~^{^+}\!\!\!\sqrt{\left(\frac{A}{\dm{}} \mp \cos 2\theta \right)^2 
\pm \sin^2 2\theta~}~;
\quad
\sin^2 2\theta_m = \sin^2 2\theta \cdot C_\pm^{-2} ~,
\label{matter}
\end{equation}
where the neutrino energy $E$ enters via the matter term 
\begin{equation}
A\equiv 2E V=\pm \frac{2\sqrt{2}G_FY\rho E}{m_n}~.
\label{Amatter}
\end{equation}
The sign of the matter term $A$ is for electron neutrinos 
(antineutrinos) traveling inside the earth positive (negative) 
and leads thus to different corrections for neutrinos ($C_+$) 
and antineutrinos ($C_-$). Matter effects modify the 2x2 vacuum 
transition probabilities
\begin{equation}
P(\nu_i\rightarrow\nu_j) = \sin^2 2\theta\cdot
\sin^2\left(\frac{\Delta m^2 L}{E}\right)\quad {\rm and} \quad 
P(\nu_{i,j}\rightarrow\nu_{i,j}) = 1-P(\nu_i\rightarrow\nu_j)~,
\label{2flavour}
\end{equation}
since in matter one must replace 
$\sin^2 2\theta\rightarrow \sin^2 2\theta_m$ and
$\dm{}\rightarrow \dm{m}$.
The matter corrections in eq.~(\ref{matter}) 
depend for a given neutrino species (i.e. given A) clearly 
on the sign of $\dm{}$ allowing thus in principle an 
extraction of this sign. One could, for example, compare 
neutrinos with antineutrinos, since the matter corrections 
go then in opposite direction inducing an asymmetry. The biggest 
matter effects (and therefore the best sensitivity to the sign
of $\dm{}$) occur when eq.~(\ref{matter}) becomes ``resonant'' for 
\begin{equation}
A= \dm{}\cos 2\theta~.
\label{optimal1}
\end{equation}
This condition can be fulfilled for a given sign of $\dm{}$ 
either for neutrinos or for antineutrinos at a specific resonance
energy, but optimization will be more complicated in a real 
experiment with an energy spectrum and other free parameters. 
Nevertheless eq.~(\ref{optimal1}) leads already to a rough 
approximation for a ``optimal mean neutrino energy''. In earth 
eq.~(\ref{optimal1}) gives for small $\theta$, i.e. for 
$\cos 2\theta=1$ roughly the relation
\begin{equation}
E_{opt} = 15~{\rm GeV}~~
\left(\frac{\dm{31}}{3.5\times 10^{-3}~{\rm eV^2}} \right) \cdot
\left(\frac{2.8~{\rm g/cm^3}}{\rho} \right) ~,
\label{optimal2}
\end{equation}
i.e. an optimal energy of about 15~GeV in the crust of the 
earth with an average effective density 
$\left\langle \rho \right\rangle \simeq 2.8~{\rm g/cm}^3$  
going down to about 10~GeV 
for paths crossing the earth at a distance of 7332~km with 
$\left\langle \rho \right\rangle \simeq 4.2~{\rm g/cm}^3$.

Our paper is organized as follows. First we develop in 
section~\ref{sec:SEC-formulae} analytic expressions for 
neutrino oscillations in matter in a suitable approximation. 
In section~\ref{sec:framework} we describe the experimental
framework on which our numerical results are based. 
Section~\ref{sec:SEC-totrates} describes the effects of 
matter on the total rates and the used statistical methods
are outlined in chapter~\ref{sec:SEC-statistics}. 
Section~\ref{sec:SEC-results} contains our results on the
sensitivity to the magnitude and sign of $\dm{31}$
and for $\sin^2 2\theta_{13} $.


\section{Analytic Description of three Neutrino Mixing in Matter 
\label{sec:SEC-formulae}} 

The mixing of three neutrinos can be described via 
\begin{equation}
|\nu_l>~ = \sum_{k=1}^3 U_{lk}|\nu_k>~, 
\hspace{1cm} l=e,\mu,\tau, 
\label{numixing}
\end{equation}
where $|\nu_l>$ corresponds to the neutrino flavour state $\nu_l$
and $|\nu_k>$ corresponds to the neutrino mass eigenstate $\nu_k$ 
with eigenvalues $m_k$, $m_k \neq m_j$, $k \neq j = 1,2,3$. $U$ is 
a $3 \times 3$ unitary leptonic mixing matrix which we parameterize 
as
\begin{equation}
\left(\begin{array}{ccc}
U_{e1}& U_{e2} & U_{e3} \\
U_{\mu 1} & U_{\mu 2} & U_{\mu 3} \\
U_{\tau 1} & U_{\tau 2} & U_{\tau 3} 
\end{array} \right)
= \left(\begin{array}{ccc} 
c_{12}c_{13} & s_{12}c_{13} & s_{13}e^{-i\delta}\\
- s_{12}c_{23} - c_{12}s_{23}s_{13}e^{i\delta} & 
 c_{12}c_{23} - s_{12}s_{23}s_{13}e^{i\delta} & s_{23}c_{13}\\
s_{12}s_{23} - c_{12}c_{23}s_{13}e^{i\delta} 
& -c_{12}s_{23} - s_{12}c_{23}s_{13}e^{i\delta}
& c_{23}c_{13}\\ 
\end{array} \right)~
\label{Umix}
\end{equation}
where $c_{ij}=\cos\theta_{ij}$ and $s_{ij}=\sin\theta_{ij}$.
The mixing matrix contains in general further CP-violation phases 
which do not enter in neutrino oscillation and are therefore
not specified \cite{BHP80}. The angles are assumed to be in the 
ranges $0 \leq \theta_{12}, \theta_{23}, 
\theta_{13} < \pi/2,~ 0 \leq \delta < 2\pi$.

\begin{figure}[htb]
\begin{center}
\epsfig{file=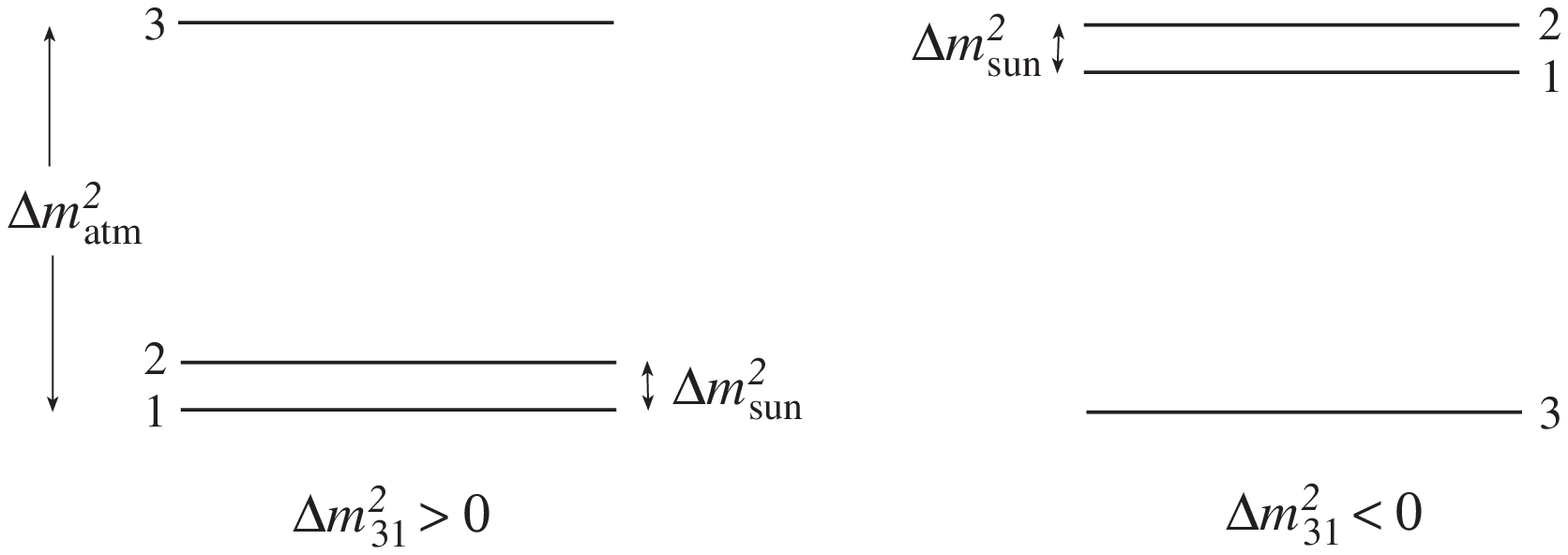,width=0.8\textwidth}
\end{center}
\mycaption{Two mass ordering schemes determined by the sign of $\dm{31}$.}
\label{fig:schemes}
\end{figure}

The squared mass splittings $\dm{}=m^2_{j}-m^2_{i} $ show
the hierarchy $\dm{21} \ll |\dm{31}|$ as discussed in more detail
below and allow for $\dm{21} \ll 10^{-4}\eV^2$, $E \geq 1\GeV$ and 
$L \leq 10^{4}$~km the  approximation where $\dm{21}=0$ in vacuum. 
We have thus two almost degenerate mass eigenvalues and for 
$\dm{31} > 0 $ ($\dm{31} < 0 $) these almost degenerate eigenvalues 
$m_1$ and $m_2$ are lighter (heavier) than $m_3$ (see 
fig.~\ref{fig:schemes}). The usual oscillation formulas depend only 
on $\dm{31}$ and it is thus impossible to discriminate between the 
two schemes. Matter effects and CP-violating effects can lift this 
degeneracy in future measurements in very long baseline experiments. 

We can assume therefore for the following discussion that there is only 
one non-vanishing quadratic mass splitting $\dm{32}=\dm{31}=\dm{}$
in vacuum. To understand this case in matter
we can write the mixing matrix eq.~(\ref{Umix}) as a sequence of 
rotations $U = R(\theta_{23})\cdot R(\theta_{13})\cdot R(\theta_{12})$
where the CP-phase $\delta$ is included in eq.~(\ref{Umix}) 
by a complex rotation in 13-subspace, i.e. by replacing 
$s_{13} \rightarrow s_{13}e^{-i\delta}$.
Alternatively the CP-phase could be included by making any of the
three rotations complex. This becomes useful when we write the
full Hamiltonian in matter with the help of $U$ in flavour 
basis\footnote{A constant common neutrino mass can be 
separated leading to an overall phase.}
\begin{equation}
{\cal H} = \frac{1}{2E} 
R(\theta_{23})\left [
R(\theta_{13})
\left(\begin{array}{ccc}
0 & 0 & 0 \\
0 & 0 & 0 \\
0 & 0 & \dm{}
\end{array} \right)
R(\theta_{13})^T
+
\left(\begin{array}{ccc}
A & 0 & 0 \\
0 & 0 & 0 \\
0 & 0 & 0
\end{array} \right)
\right ]R(\theta_{23})^T~.
\label{Hamiltonian}
\end{equation}
Here use has been made of the fact $R(\theta_{12})$, which operates
in 12-subspace, drops out in the limit where $\dm{21}=0$. Since we 
could use a convention where the CP-phase would be defined by a 
complex rotation in 12-subspace we can immediately see that the 
CP-phase drops out as well. Furthermore $R(\theta_{23})$ operates 
in the 23-subspace and commutes therefore with the matter term. 
We see that matter effects are confined in 
eq.~(\ref{Hamiltonian}) inside the square brackets, which implies 
that matter affects in the $\dm{21}=0$ limit only the 13-subspace. 
We can readily use the standard two neutrino parameter 
mapping given in eqs.~(\ref{matter}) for the 13-subspace and 
obtain in matter $\dm{31,m}= \dm{}C_\pm$ and 
$\sin^2 2\theta_m =\sin^2 2\theta/C_\pm^2$, where $\dm{}\equiv\dm{31}$ 
enters into $C_+$ defined in eq.~\ref{matter}. Note, however, 
that $\dm{32}$ and $\dm{21}$ become in matter in this way also 
energy dependent, namely
\begin{equation}
\dm{32,m} = \dm{} C_\pm^{\prime} = \frac{\dm{}~(C_\pm+1)+ A}{2}~; \quad
\dm{21,m} = \dm{} C_\pm^{\prime\prime}= \frac{\dm{}~(C_\pm-1)- A}{2}~.
\label{Dmijmatter}
\end{equation}
An important consequence is that 
the mass degeneracy between the first two eigenvalues is 
destroyed in matter. Thus even if we can approximate $\dm{21}=0$ 
in vacuum we need for oscillation in matter the full three 
neutrino oscillation formulae. The only simplification left is that
one can set in the mixing matrix $\theta_{12}=0$ and $\delta=0$.

These matter induced parameter mappings must now be inserted 
into the oscillation formulae for three neutrinos. 
Defining as shorthands $D_{ab}= e^{-iE_at}~\delta_{ab}$,
$J_{ij}^{lm} := U_{li}U_{lj}^*U^*_{mi}U_{mj}$ and 
$\Delta_{ij} := \dm{ij} L / 4E$ 
the transition probabilities $P(\nu_l\rightarrow\nu_m)$ 
from flavour $l$ to flavour $m$ in vacuum can be written as
\begin{equation}
P(\nu_l\rightarrow\nu_m) 
= \left| \left\langle l | UDU^+ | m \right\rangle \right |^2
= \delta_{lm} - 4\sum_{i>j} \mathrm{Re} J_{ij}^{lm}~\sin^2\Delta_{ij} 
- 2\sum_{i>j} \mathrm{Im} J_{ij}^{lm}~\sin 2\Delta_{ij}~.
\label{Plm}
\end{equation}
As described above we can set in matter $\theta_{12}=0$ and 
$\delta=0$ resulting in
\globallabel{Ps}
\begin{align}
P(\nu_\mu\leftrightarrow\nu_e) = &  
~\sin^2 \theta_{23}\sin^2 2\theta_{13} \sin^2(\Delta_{31})~, \mytag \\
P(\nu_\mu\leftrightarrow\nu_\mu) = &   
~1 - \sin^2 2\theta_{23} \sin^2\theta_{13} \sin^2(\Delta_{21})
 - \sin^4 \theta_{23} \sin^2 2\theta_{13} \sin^2(\Delta_{31}) \nonumber \\
 &~ - \sin^2 2\theta_{23} \cos^2\theta_{13} \sin^2(\Delta_{32}) ~,\mytag\\
P(\nu_\mu\leftrightarrow\nu_\tau) = &  
~\sin^2 2\theta_{23}\left [     
\sin^2\theta_{13} \sin^2(\Delta_{21})
-\frac{1}{4}\sin^2 2\theta_{13} \sin^2(\Delta_{31}) \right . \nonumber \\
 &~ \left . 
+ \cos^2\theta_{13} \sin^2(\Delta_{32})\mytag
\right ]~.
\end{align}
To describe oscillation in matter we must insert into these 
formulae the parameter mappings discussed above, namely
the shifted mass eigenvalues ($\dm{31,m}=\dm{}C_\pm$ and 
eqs.~(\ref{Dmijmatter})) and the shifted 13-mixing angle, 
$\sin^2 2\theta_{13,m} = \sin^2 2\theta_{13}/C_\pm^2$.
Note that the relations for antineutrinos are only formally 
identical due to the 
opposite sign of the matter term $A$ in eq.~(\ref{matter}).
In vacuum the relations~(\ref{Ps}a)--(\ref{Ps}c) could be further 
simplified since the tight CHOOZ results \cite{CHOOZ99} can be
taken into account. For typically allowed $\dm{31}$ values one has 
$\sin^2 2\theta_{13} \ltap 0.1$, while this upper limit 
is less stringent for $1.0\times 10^{-3}\eV^2 
\leq  |\dm{31}| < 3.0\times 10^{-3}\eV^2$ where 
values of $\sin^2 2\theta_{13} \simeq 0.2$ are allowed.
Note however that the effective $\sin^2 2\theta_{13}$
in matter can be much bigger when the resonance condition is 
fulfilled such that further simplifications of 
eqs.~(\ref{Ps}a)--(\ref{Ps}c) are potentially dangerous.


\section{Experimental Framework
\label{sec:framework}}
 
We use for our study a standard earth matter density profile
\cite{Stacey77,PREM81} and we will treat neutrino oscillations 
in earth by a constant density approximation along each 
neutrino path. The varying matter profile of the earth implies
that the average density is still path dependent. It has been
shown recently \cite{Shrock99,MartinTommy} that this approximation 
works very well, while a constant density for all paths ignores 
even the mean density variations in the earth and has sizable 
errors\footnote{Note also that care should be taken about the 
core of the earth. Either one should avoid paths passing the 
core or one should take core effects into account. Specific 
core effects become however small when the oscillation 
wavelength becomes bigger than the size of the core.}.
We avoid the core of the earth and take in our study distances 
up to 7332~km. 
We assume furthermore a neutrino factory which can operate 
with $\mu^-$ or $\mu^+$ beams leading to $\nu_\mu + \bar \nu_e$ 
or $\bar \nu_\mu + \nu_e$ neutrino beams resulting from
$N_{\mu^{\pm}}$ muon decays. The detector with $\NKT$ kilotons 
is assumed to be able to measure muons above an assumed 
threshold of $5$~GeV \cite{FNALworkshopA} with an efficiency 
$\epsilon_\mu$. The muons may come from the dominant 
$\ruu$ disappearance  channel or from the subdominant $\reu$ 
appearance channel with or without charge identification. 
Muons coming from the subdominant $\rut$ or $\ret$ channels with 
the $\tau$ decaying subsequently to a muon should not be a 
problem, since they can be separated by kinematical means without 
loosing too much muon detection efficiency \cite{BARGER99b}.
The inclusion of various other potential backgrounds and detector 
properties will have some influence on the analysis, but with 
current understanding this would only result in moderate 
corrections. More detailed simulations of backgrounds are still in 
progress \cite{FNALworkshopA,FNALworkshopB,FNALworkshopC,FNALworkshopD}. 
Most limitations due to backgrounds and statistics can however
be overcome by increasing the number of muon decays 
and/or the size of the detector. We do not include 
a detailed simulation of background effects at the moment.

For three neutrinos there is only room for two independent quadratic 
mass splittings while there are three different results for 
oscillation. Among these the LSND evidence \cite{LSNDev} for 
oscillation is most controversial and we omit this result therefore 
in our analysis, while we take the atmospheric result (mostly from 
SuperKamiokande \cite{SKev}) and the solar neutrino deficit (mostly 
from GALLEX \cite{GALLEXev}) as evidence for oscillation. 
These results can be studied in a simplified 2x2 oscillation picture where 
the mixing matrix contains only one mixing angle $\theta$ with 
the two flavour transition probabilities in vacuum given in 
eq.~(\ref{2flavour}). The dominant atmospheric 
$\nu_{\mu} \leftrightarrow \nu_{\tau}$ oscillations imply in 
this picture a mass splitting \cite{atmdm2}
\begin{equation}
10^{-3}\eV^2\ltap |\dm{31}| \ltap
8.0\times 10^{-3}\eV^2~,
\label{Dm2atm}
\end{equation}
while the solar neutrino deficit leads to different solutions 
 \cite{solardm2} for $\dm{21}$, all with $|\dm{21}| \ll |\dm{31}|$.
These are the solar vacuum oscillation (VO) solution and the large 
and small mixing angle (LMA and SMA) solar MSW solutions. The above 
assignment implies thus that $\dm{31}$ dominates the atmospheric  
$\rut$ oscillation, while $\dm{21}$ is most relevant for 
the VO, SMA~MSW and LMA~MSW solution of the solar neutrino problem. 
We will see that our study depends only on $\dm{31}$ such that  in a
very good approximation $\dm{21}$ can be set to zero. We use
therefore as default initial parameters of our study
\begin{equation}
| \dm{31} |  = 3.5 \cdot 10^{-3} ~;\quad
| \dm{21} |  = 0 ~;\quad
\sin^2 2 \theta_{23} = 1.0
\end{equation}
and we will discuss variations of these parameters within the 
allowed ranges.

The event rates of our results depend only on the combination 
$N_{\mu^{\pm}} \NKT \epsilon_\mu$ and we take for both polarities 
$N_{\mu^\pm} \NKT \epsilon_\mu= 2 \cdot 10^{21}$.
This corresponds for example to a setup with 
$N_{\mu^+} = 2 \cdot 10^{20} \mu^{+}$ decays, 
$N_{\mu^-} = 2 \cdot 10^{20} \mu^{-}$ decays,
an $\NKT=10$~kt iron detector and a muon detection efficiency 
$\epsilon_\mu=1$. Unless otherwise mentioned we use a 
baseline of $L=7332$~km.

Our analysis is performed at the level of total and differential 
event rates. We include therefore the charged current neutrino 
cross sections per nucleon in the detector and the normalized 
$\nu_\mu + \bar\nu_e$ and $\bar\nu_\mu + \nu_e$ beam spectra 
$g_{\nu_i}$ of the neutrino factory \cite{GEER98}
\globallabel{Xsection}
\begin{align}
\sigma_{\nu_\mu}(E) &= 
0.67\cdot 10^{-38}E\,\text{cm}^2/\text{GeV}~, &
\sigma_{\bar{\nu}_\mu}(E) &= 
0.34\cdot 10^{-38}E\,\text{cm}^2/\text{GeV}~, \mytag\\
g_{\nu_e}(x) &= g_{\bar{\nu}_e}(x) = 12x^2(1-x)~, &
g_{\nu_\mu}(x) &= g_{\bar{\nu}_\mu}(x) = 2x^2(3-2x)~, \mytag
\end{align}
where $x=E/E_\mu$. For a given number of useful muon decays 
$N_{\mu^\pm}$, detector size $\NKT$, efficiency $\epsilon_\mu$
one obtains for the channel "i" (where i stands for a proper
index uniquely given by the polarity of the muon beam and the considered
oscillation channel) the contributions to the differential event 
rates
\begin{equation}
\frac{dn_i}{dE} = \underbrace{ \left [ N_{\mu^i} \NKT \epsilon_\mu~
\frac{10^9N_A}{m^2_\mu\pi} \right ]~}_{normalization}
 \underbrace{\left [ \frac{E_\mu}{L^2}~
g_i(E/E_\mu) \sigma_i(E) \right ]~}_{flux}
 \underbrace{\left [ P_i(E)\frac{}{} \right ]~}_{oscillation}
\label{differential}
\end{equation}
where $10^9N_A$ is the number of nucleons per kiloton in the detector
and $P_i$ stands for the relevant oscillation probability in matter
as discussed above. In cases where different channels contribute
the total differential rates are given by the sum of all individual 
terms. Total rates are obtained by integrating these differential
rates from the threshold at $5$~GeV to the maximal possible neutrino 
energy $E_\mu$.

To understand the event rates we look at the three terms in square 
brackets on the {\sl rhs.} of eqs.~(\ref{differential}). 
The first term contains overall factors which are constant in 
energy, which is only important for the proper event rate 
normalization. The second term is simply the product of flux 
times detection cross section which has to be folded with the 
third term, the oscillation probability in matter.
The second term grows initially like $E^3$ and reaches
a maximum before it goes to zero at $E=E_\mu$. 
An important feature of the second term is that the low 
energy part is not changed when the muon energy is increased.
This implies in a good approximation that the increase in 
the total event rates by increasing $E_\mu$ comes essentially
from the high energy tail of the spectrum only. This has 
also the important consequence that experimental results 
for a lower muon beam energy $E_\mu^\prime < E_\mu$ can in principle 
be obtained to a good approximation at higher energies by 
simply removing events with energies above $E_\mu^\prime$.


\section{Total Rates
\label{sec:SEC-totrates}} 

The analytic description above allows now an understanding of 
the event rates which we calculated numerically 
and which are presented below in fig.~\ref{fig:ratessep}.
To understand first the total disappearance rates without matter 
one must look at the folding of the vacuum oscillation probabilities 
with the flux factor in eq.~(\ref{differential}). The oscillatory 
terms in the oscillation probabilities eqs.~(\ref{Plm}), namely the 
$\sin^2 \Delta_{ij}$ factors, depend then for fixed $\dm{31}=\dm{}$ 
only on $E/L$. The oscillatory behavior vanishes for 
$E/L \rightarrow \infty$ and the first maximum of the vacuum 
oscillation occurs for given $L$ at an energy $E_1=2\dm{}L/\pi$. 
This means that there will essentially be no effects of oscillation 
in the total disappearance rates when $E_1$ is much smaller than the 
neutrino energy $E$. Since the event rates are dominated by the 
maximum of the flux factor close to $E_\mu$ we can set $E\simeq E_\mu$ 
and find no oscillation effects when $E_\mu$ is much larger than $E_1$. 
The overall rates will however still decrease with the muon energy $E_\mu$ 
according to the cross sections and fluxes approximately like $E_\mu^2$ 
as long as $E_\mu$ is sufficiently larger than the muon threshold 
energy of $5$~GeV. The first effects of vacuum oscillation show up 
for lower beam energies when $E\simeq E_\mu$ is decreased such 
that $E_\mu= {\cal O}(E_1)$. The oscillation becomes maximal
for $E_\mu \simeq E_1$ while it vanishes again for $E_\mu\simeq E_1/2$.
The effects of oscillation can be seen in the total rates as a
dip in the overall $E_\mu^2$ scaling. For lower beam energies 
there are in principle further dips due to all other oscillation
maxima, but this is on our case below the threshold energy of $5$~GeV.

In order to understand how matter effects influence the  
disappearance rates one must compare the matter 
``resonance energy'' $E_{opt}$ in eq.~(\ref{optimal2}) at 
roughly $10-15$~GeV with $E_1$ and $E_\mu\simeq E$.
At neutrino energies which are not close to this resonance 
energy the mixing $\sin^2 2\theta_{13,m}$ approaches quickly 
its value in vacuum, which is small. One can see 
immediately from the Hamiltonian in eq.~(\ref{Hamiltonian}) 
that the dominant $\rut$ oscillation decouples 
in the limit where $\sin^2 2\theta_{13}=0$ from the 
matter effects coming from the first generation such that
it is clear that matter effects should be rather localized
around the resonance region\footnote{Note that the mass splittings
$\dm{ij}$ behave non-trivial in the high energy limit such
that the asymptotic properties of eq.~(\ref{Ps}b) have to
be evaluated carefully to reach the same conclusion.}.

The total appearance rates are given for $E_\mu \gg E_1$ by 
the asymptotic behavior of eq.~(\ref{Ps}a) weighted with the 
flux. The increase in the rates with $E_\mu$ comes again 
predominantly from the increased flux at energies close to 
$E_\mu$ such that we need to look again in a good approximation
at the asymptotic behaviour of 
$E_\mu^2 P(\nu_\mu\leftrightarrow\nu_e)$. Expanding 
$E_\mu^2 P(\nu_\mu\leftrightarrow\nu_e)$ in powers of $1/E$, 
and using the fact that $V\cdot L$ is numerically small, one finds 
for the asymptotic appearance rates for small $\sin^2 2\theta_{13}$
\begin{equation}
n \propto \sin^2 \theta_{23} \sin^2 2\theta_{13}
(\dm{})^2 L^2 \left(1-\frac{(\dm{})^2L^4}{3E_\mu^2}
\right ) +\frac{4(\dm{})^3 L^4V}{3E_\mu}~.
\label{expandP}
\end{equation}
The result shows that the rates approach for $E_\mu \gg E_1$
in vacuum a constant which depends only on even powers of 
$\dm{}$. The leading corrections to this constant fall like 
$1/E_\mu^2$. Turning on matter effects leads in eq.~(\ref{expandP}) 
to the third term which falls only like $1/E_\mu$ and which depends 
on $(\dm{})^3$, i.e. on the sign of $\dm{31}$. This induces in
our case a matter dependent splitting in the appearance rates 
which is less localized in energy than the matter effects in
the disappearance channels. There are thus significant effects 
even for beam energies which are somewhat away from the 
resonance energy. Note, however, that the discussion of 
effects is much more complicated for muon energies of the order
or smaller than $E_1$.

\begin{figure}[htb!]
\begin{center}
\epsfig{file=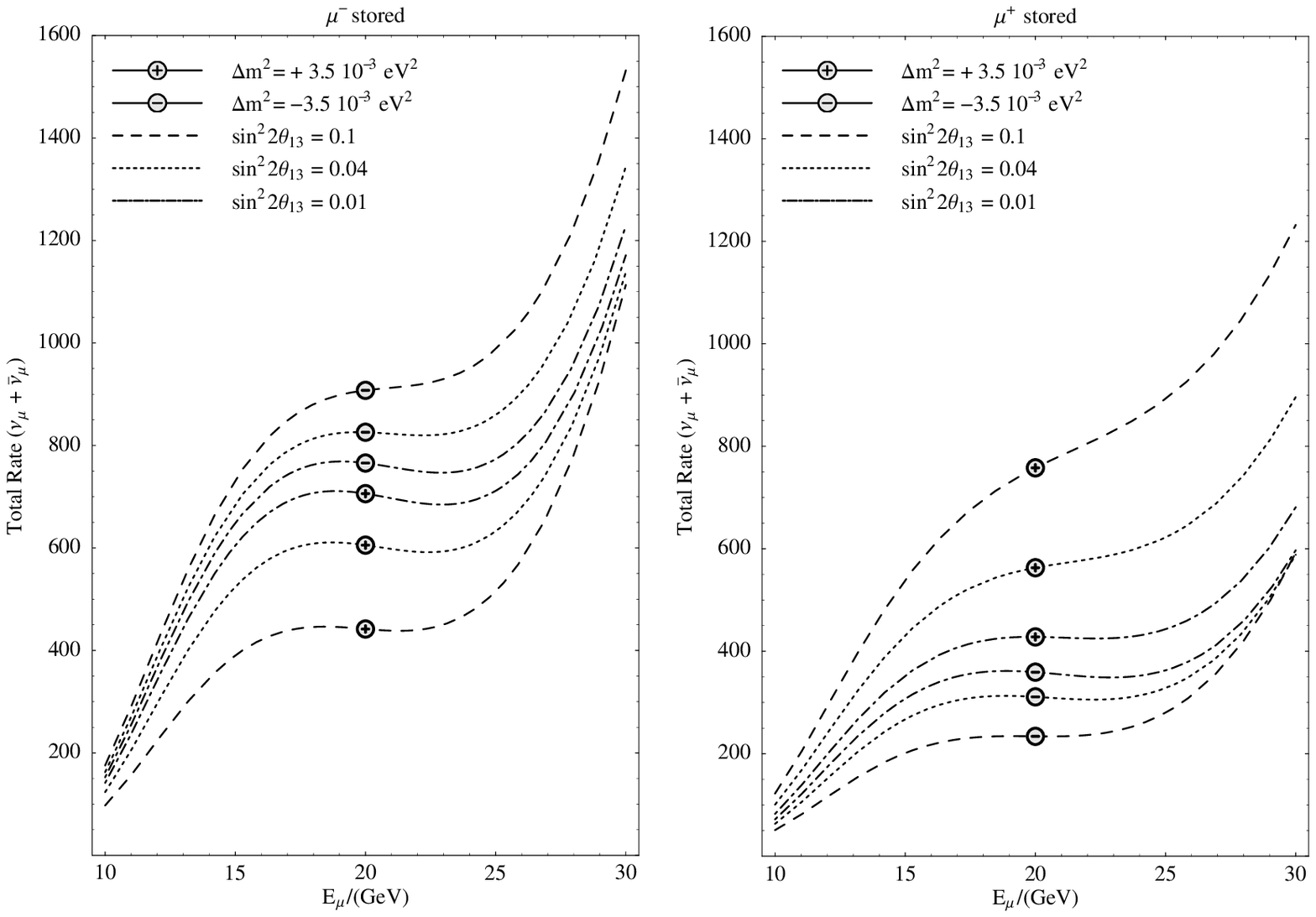,width=\textwidth}
\end{center}
\mycaption{Total muon rates $\nu_\mu + \bar{\nu}_\mu$ (i.e. $\ruu$ 
combined with $\reub$ and $\ruub$ combined with $\reu$) as a function 
of the muon beam energy for a baseline of 7332~km, 
$|\dm{31}|=3.5 \cdot 10^{-3}~{\rm eV}^2$ and 
$N_{\mu^\pm}\NKT\epsilon_\mu= 2\cdot10^{21}$ which corresponds typically 
to a $10$~kt-year. 
The different line types stand for different mixing angles $\theta_{13}$
(see labels). $\oplus$ and $\ominus$ indicate positive and negative 
mass squared difference.}
\label{fig:rates35}
\end{figure}

\begin{figure}[htb!]
\begin{center}
\epsfig{file=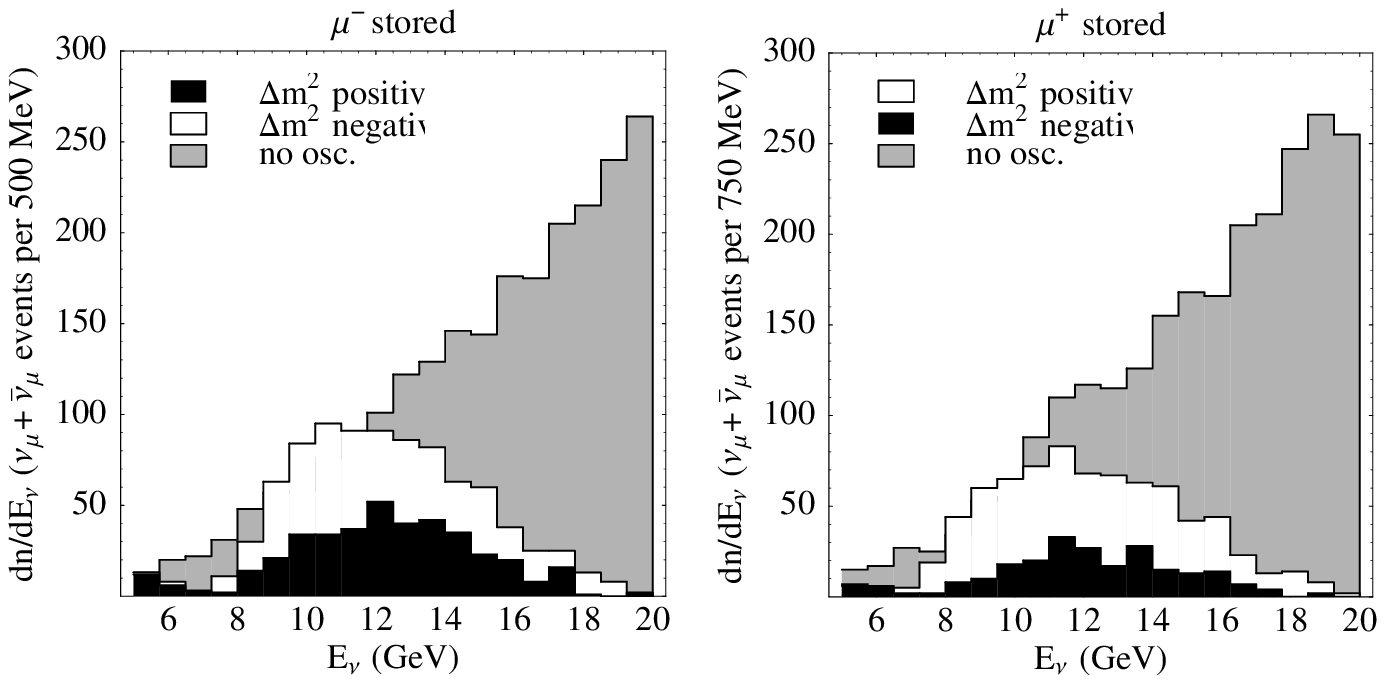,width=14cm}
\end{center}
\mycaption{Spectral distribution of the muon rates $\nu_\mu + \bar{\nu}_\mu$
according to fig.~\ref{fig:rates35} for $\sin^2 2\theta_{13} = 0.1$. The left 
plot was done for negative stored muons, the right plot for positive muons. 
We use 20 energy bins over the range from $5 \GeV$ to $20 \GeV$.}
\label{fig:spectrum}
\end{figure}

This understanding of matter effects can now be compared with the 
results of our full numerical calculations. The following figures 
for the total rates in the different oscillation channels assume 
$N_{\mu^\pm}\NKT\epsilon_\mu= 2\cdot10^{21}$, a fixed 
baseline of 7332~km, $|\dm{31}| = 3.5\cdot 10^{-3} \eV^2$, 
$\sin^2 2\theta_{23} = 1$ and different values of 
$\sin^2 2\theta_{13}$. If the muon beam energy is fixed then
we use always $E_\mu=20 \GeV$.
\begin{figure}[htb!]
\begin{center}
\epsfig{file=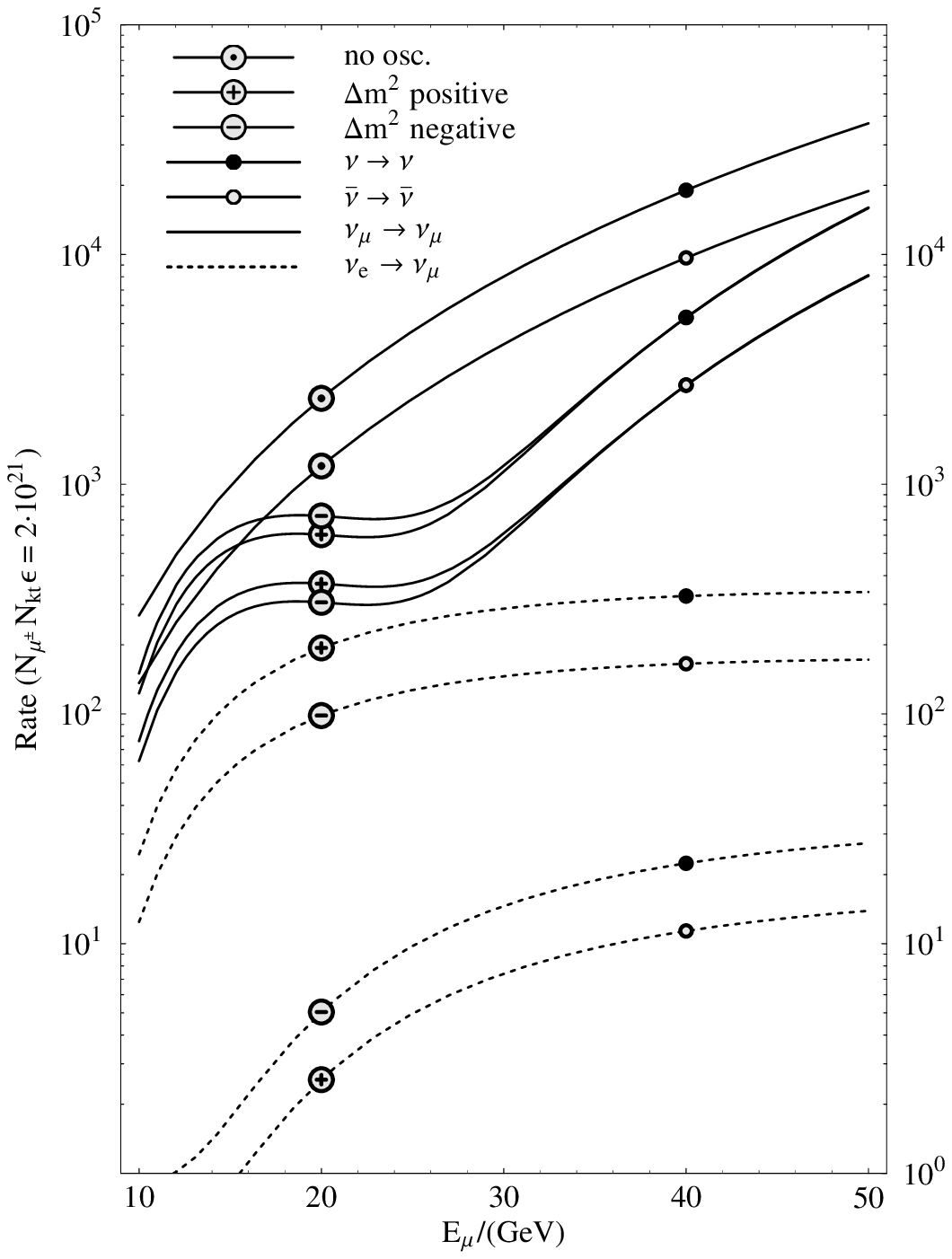,width=8cm}
\end{center}
\mycaption{Muon appearance rates $\reu$ and $\reub$ (dotted) 
and muon disappearance rates $\ruu$ and $\ruub$ (solid) as a 
function of the muon beam energy for a baseline of 7332~km, 
$|\dm{31}|=3.5\cdot 10^{-3}~{\rm eV}^2$ and $\sin^2 2\theta_{13} = 0.1$. 
$\oplus$ and $\ominus$ indicate positive and negative mass squared 
difference, $\odot$ indicates no oscillation rates. $\bullet$ and $\circ$ 
label the neutrino and the antineutrino channels.}
\label{fig:ratessep}
\end{figure}
Fig.~\ref{fig:ratessep} shows the $\reu$ and $\reub$ total appearance 
rates (dotted lines) as well as the $\ruu$ and $\ruub$ disappearance 
rates (solid lines) on a logarithmic scale for maximal 
$\sin^2 2\theta_{13} = 0.1$. These rates are in agreement with the 
results obtained recently by Barger et. al \cite{BARGER99b}. 
The lines appear in pairs of neutrino ($\bullet$) and antineutrino 
($\circ$) channels which differ roughly by a factor two coming from 
the cross sections eqs.~(\ref{Xsection}). Fig.~\ref{fig:ratessep} shows
the cases with positive $\dm{31}$ 
($\oplus$), negative $\dm{31}$ ($\ominus$) and no oscillation 
($\odot$). The asymmetry between $\oplus$ and $\ominus$ in the appearance 
rates, which has been used to extract the sign of $\dm{31}$ at smaller 
baseline \cite{BARGER99b,FLPR}, 
is clearly visible. Note that there are also comparable matter effects 
in the disappearance channels at these large baselines, which appear
only to be smaller due to the logarithmic scale. Fig.~\ref{fig:ratessep} 
shows clearly the important feature that the matter effects in the
combined appearance and disappearance muon rates go for a neutrino 
factory with either $\nu_\mu+\bar\nu_e$ or $\bar\nu_\mu+\nu_e$ beams 
in the same direction. This opens an interesting possibility if muon 
charge identification is not available to separate the appearance 
channels (wrong sign muon events) from the disappearance channels 
(right sign muon events). It is then still possible to measure all 
muons, i.e. muons with both charges, or in other words the combination 
of the two oscillation channels, where matter effects add up. The 
resulting total muon rates of the combined channels are shown for 
both $\mu^-$ and $\mu^+$ beams in fig.~\ref{fig:rates35} on a 
linear scale for $\mu^-$ and $\mu^+$ muon beams. The figure shows 
the sizable matter induced splittings in the total muon-neutrino rates 
($\nu_\mu + \bar{\nu}_\mu$) for the mixing angles 
$\sin^2 2\theta_{13} = (0.1, 0.04, 0.01)$.
The difference between the two signs of $\dm{31}$ influences 
the total rates up to a factor two or more. The effect comes partly from 
the appearance rates and partly from the disappearance rates, but
always from the resonant channel. This is for positive $\dm{31}$
the neutrino component of the beam and for negative $\dm{31}$
the antineutrino component. For $\mu^+$ beams the interplay of the 
two channels is as follows\footnote{For $\mu^-$ beams holds the same 
for a reversed sign of $\dm{31}$.}:
For positive $\dm{31}$ the $\reu$ appearance channel is resonant 
leading to an enhancement of the combined rate, while matter corrections
are moderate for the $\ruub$ disappearance channel. For negative $\dm{31}$,
on the other side, the disappearance channel $\ruub$ shows a resonant 
transition of $\nu_\mu$ to $\nu_e$ resulting in a suppression of the 
total $\nu_\mu + \bar{\nu}_\mu$ rate, while the $\reu$ appearance rates
are mostly unchanged. The combination of appearance and disappearance 
channels amplifies thus always the signal. This shows that 
an analysis of the combined appearance and disappearance channels
is interesting and it may be especially important if the separation 
of right and wrong sign muon events is not available.
Figure~\ref{fig:spectrum} shows the corresponding differential 
event rate spectrum of the  $\nu_\mu + \bar{\nu}_\mu$ induced 
muon rate over twenty energy bins for a mixing angle 
$\sin^2 2\theta_{13} = 0.1$ (CHOOZ bound).
We find that the different matter effects have no specific spectral
features which could be used as a basis for cuts in order to amplify 
the signal relative to the oscillation signal. We will fit therefore 
in the following differential event rate spectra of data simulated 
with certain initial parameters and with statistical noise added. 
We will show how well these parameters can be re-extracted in an appropriate
statistical way. We provide therefore in the next section information
about the used statistical procedures and we will study then in more 
detail the statistical significance of the matter effects in the 
appearance channels, disappearance channels and combinations thereof. 
We focus on the potential to extract $\theta_{13}$, the sign of 
$\dm{31}$ and to test matter effects.


\section{Statistical Methods and Data Evaluation
\label{sec:SEC-statistics}} 

We numerically simulate data for a given parameter set 
($E_\mu$, $\dm{31}$, $\sin^2 2\theta_{23}$, $\sin^2 2\theta_{13}$) 
and add Poisson distributed fluctuations to the energy bins. 
The analysis of these ``data'' is done according to  the procedure 
proposed by the Particle Data Book \cite{PDG99} for Poisson 
distributed quantities. Confidence levels (CL) are calculated by using
\begin{equation}
\label{chi}
\chi^2 = \sum_{i=1}^{n} \left( 
2\, \left[n_i^\mathrm{th}-n_i^\mathrm{obs}\right] +2 n_i^\mathrm{obs} 
\log\frac{n_i^\mathrm{obs}}{n_i^\mathrm{th}} \right) \quad . 
\end{equation}
We assume symmetric operation of the neutrino factory in both polarities,
i.e. $\mu^{+}$ and $ \mu^{-} $ beams together with a detector such that
$N_{\mu^+}\NKT\epsilon_\mu = N_{\mu^-}\NKT\epsilon_\mu = 2 \cdot 10^{21}$.
The total $\chi^2$ is given by
\begin{equation}
\chi^2 = \chi^2_\mathrm{\mu^+} + \chi^2_\mathrm{\mu^-}~.
\end{equation}
If there is no charge identification then 
$n_i = (n_{\mu^+})_i + (n_{\mu^-})_i$ is the total 
(indistinguishable) number of $\nu_\mu$ and $\bar{\nu}_\mu$ induced 
muon events in the i-th energy bin and {\em obs} and {\em th} 
label the observed (i.e. simulated) and theoretical predicted muon event
rates. With charge separation we calculate $ \chi ^2_\mathrm{\mu^{\pm}}$ 
separately for neutrinos and antineutrinos and use the sum
\begin{equation}
\chi ^2 = \chi ^2_\mathrm{\mu^- ,\nu}+\chi ^2_\mathrm{\mu^- ,
\bar{\nu}}+\chi ^2_\mathrm{\mu^+ ,\nu}+\chi ^2_\mathrm{\mu^+ ,\bar{\nu}}~.
\end{equation}
Next $ \chi ^2 $ is minimized as usual with respect to the parameters shown in
the plots. We subtract the minimum value of $ \chi ^2 $ and define
confidence levels according to the values of 
\begin{equation}
\Delta \chi ^2 = \chi ^2 - \chi ^2_{min}
\end{equation}
This $ \Delta \chi ^2 $ obeys a $\chi ^2$ distribution  
for $k$ degrees of freedom (i.e. the number of parameters fitted). 
The value of $ \Delta \chi ^2_{CL} $ which corresponds to a given 
confidence level CL is given by:
\begin{equation}
CL = \int _{0}^{ \Delta \chi ^2_{CL}}
\frac{x^{\frac{k}{2}-1} e^{- \frac{x}{2}}}{2^{\frac{k}{2}} \Gamma  ( \frac{k}{2} )} \quad dx
\end{equation}
For two degrees of freedom $(1\sigma,2\sigma,3\sigma) $ corresponds to 
$ CL=(68.3\% , 95.5\% , 99.7\%) $ and to 
$\Delta \chi ^2_{CL}=(2.3,6.2,11.8)$, as usual.

We use in our analysis for the interval between 5 GeV and 20 GeV
a total of 20 bins. We checked that the statistical 
significance of the results does not change much by changing the
number of bins, as long as one has enough bins to describe the 
spectral information sufficiently. Finer binning does thus not 
improve the significance due to the statistical limitation 
by fluctuations. In this sense it is best to use around 20 energy
bins which should not be a problem with the usual energy resolution
of detectors.


\section{Results
\label{sec:SEC-results}} 

As outlined above, we simulate a measurement with certain input
parameters including statistical fluctuations for a given sign of 
$\dm{31}$. Then we try to fit this dataset, i.e. we re-extract 
the input parameters both with the right and the wrong sign of 
$\dm{31}$. Instead of a global fit to all parameters one can 
first fit the magnitude of $|\dm{31}|$ and $\sin^2 2\theta_{23}$ 
with good precision (a few percent \cite{BARGER99}) by analyzing 
the muon spectrum for the dominant $\ruu$ and $\ruub$ disappearance 
oscillation\footnote{One could combine the differential 
event rates coming from the $\mu^+$ and $\mu^-$ beams, weighted
to correct for the cross section differences, such that the 
matter dependence is almost removed. This allows a determination 
of $|\dm{31}|$ and $\sin^2 2\theta_{23}$ which is essentially 
independent of the sign of $\dm{31}$ and $\sin^2 2\theta_{13}$.}. 
The following plots assume that this analysis is done and 
we analyze then with $\sin^2 2\theta_{23}$ 
fixed the matter induced effects and extract $\sin^2 2\theta_{13}$
and the sign of $\dm{31}$. We will see that this works very 
well if the mixing angle $\theta_{13}$ is not too small. The 
value of $\Delta \chi^2$ for the wrong sign of $\dm{31}$ will 
tell us the confidence level at which we can reject the 
incorrect sign of $\dm{31}$. For the right sign we can 
calculate confidence level contour lines from which we can read 
off the sensitivity for the determination of the relevant 
oscillation parameters. The $\Delta \chi ^2 $ values printed 
in the following figs.~\ref{fig:summe} - \ref{fig:combined} 
next to the CL contour lines are for the
wrong sign of $\dm{31}$ relative to the best fit with the 
correct sign of $\dm{31}$. This are in figure \ref{fig:summe}
the values of the global minima with the wrong sign. In the
other two figures we used the information on $|\dm{31}|$ from
the fit in the $\sin^2 2\theta_{23}$-$|\dm{31}|$ plane to 
reject $\dm{31}$ values which are inconsistent with the
$|\dm{31}|$ value obtained before. We therefore restrict the fit
for the wrong sign of $\dm{31}$ to the local minimum of
$\chi^2$ in the neighborhood of the best fit of $|\dm{31}|$.
\begin{figure}[htb!]
\begin{center}
\epsfig{file=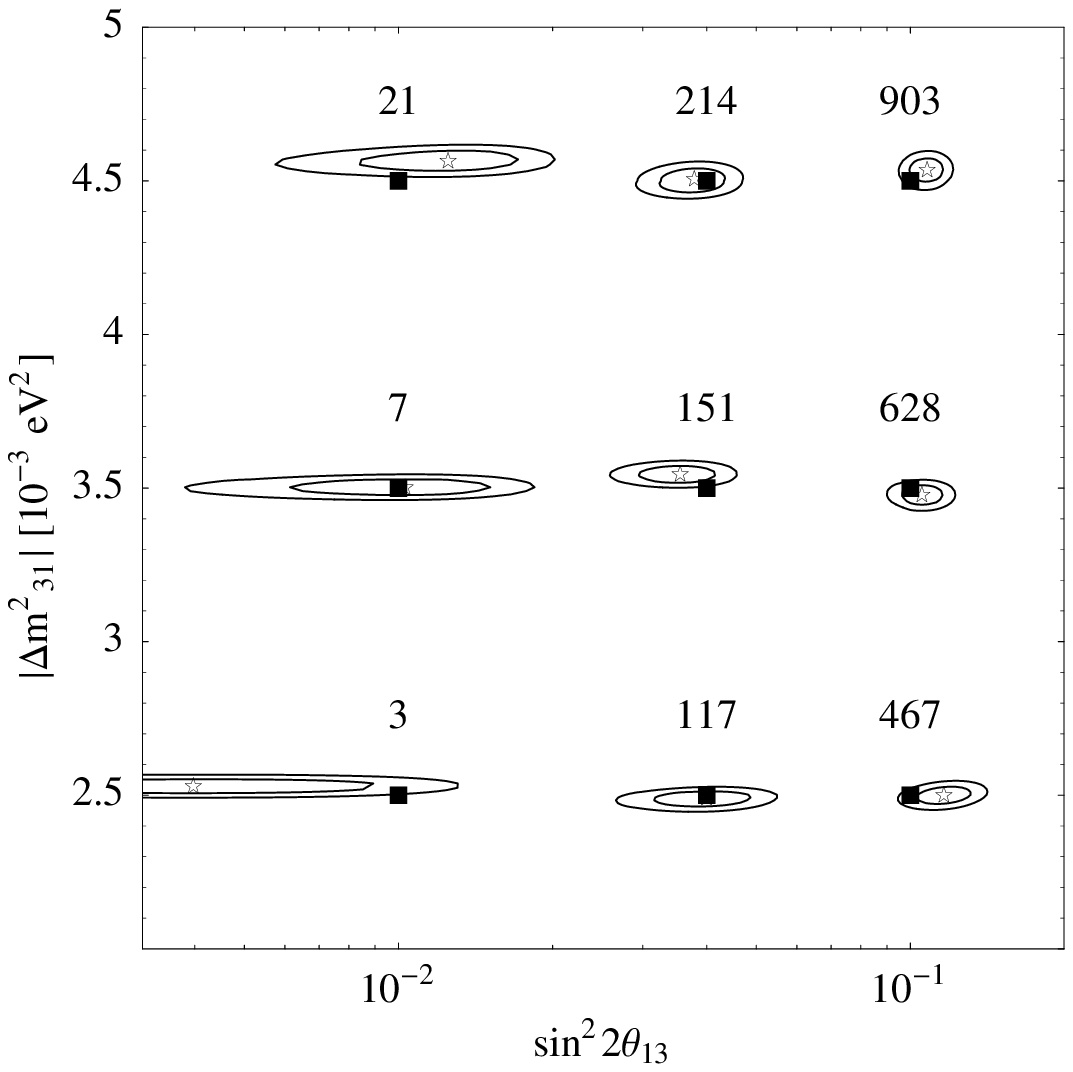,width=10cm}
\end{center}
\mycaption{Fit to the muon neutrino spectrum without charge
identification (i.e. the sum of unoscillated ($\ruu $) and oscillated
($\reub $) muon neutrinos) for $\dm{31}>0$ at a baseline of 7332~km. 
Shown are the $1\sigma$ and  $2\sigma $ contours. The rectangles denote 
the parameter pair for which the data were generated and the stars denote 
the obtained best fit. The numbers printed next to each case are the 
values of $\chi^2$ per 2~d.o.f for the best fit with the wrong sign of 
$ \dm{31} $.}
\label{fig:summe}
\end{figure}

\begin{figure}[htb]
\begin{center}
\epsfig{file=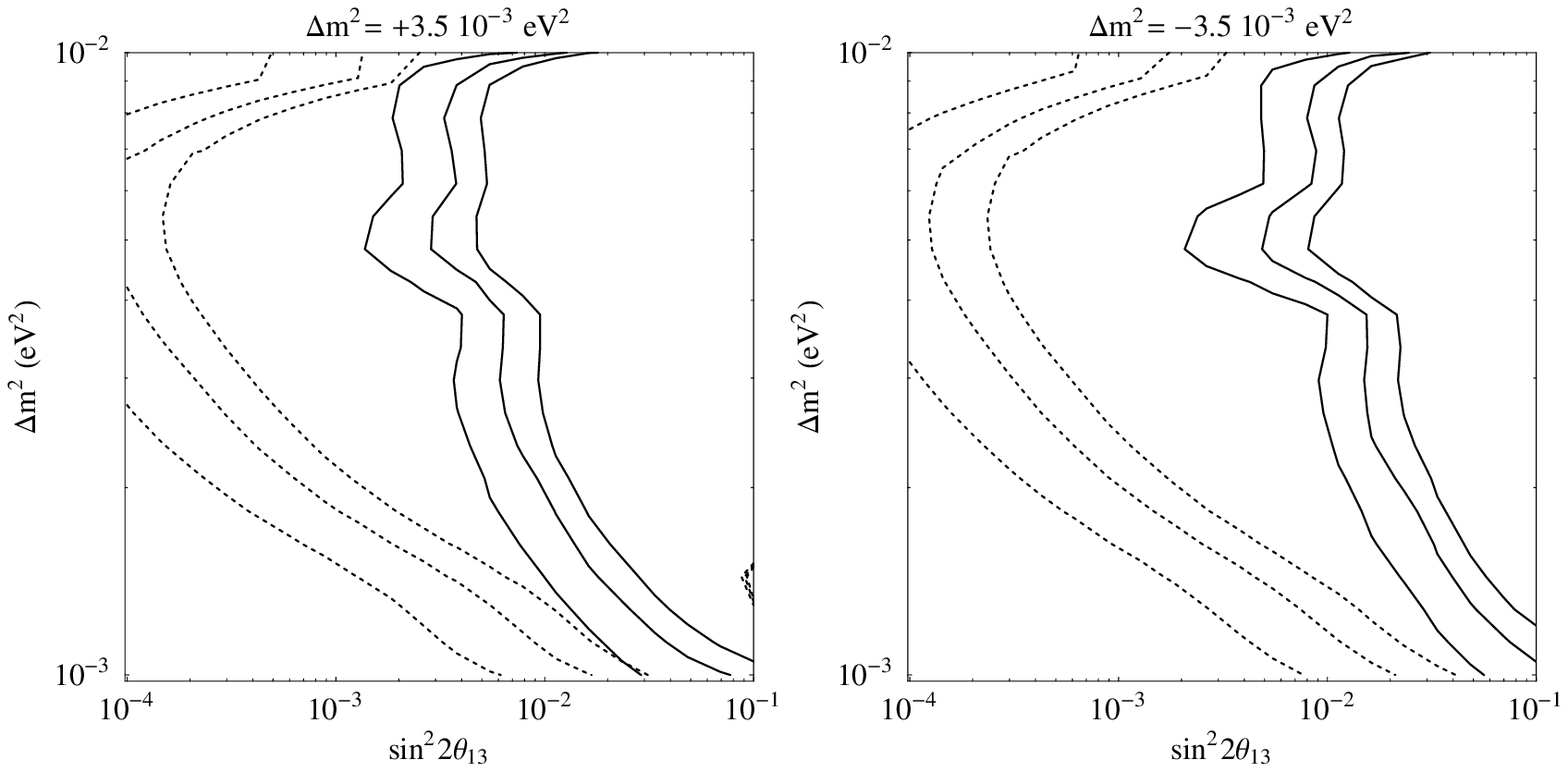,width=\textwidth}
\end{center}
\mycaption{Sensitivity to the sign of $\dm{}$. Shown are the  
$1\sigma$, $2\sigma$ and $3\sigma$ contour lines of a fit with the 
wrong sign of $\dm{}$. The dotted lines are based on spectral 
data from the appearance channel only. The solid lines use the combined 
data from the appearance and disappearance channels without separation
of right and wrong sign muons. (7332~km, 
$|\dm{31}|=3.5 \cdot 10^{-3}~{\rm eV}^2$ and 
$N_{\mu^\pm}\NKT\epsilon_\mu= 2\cdot10^{21}$).}
\label{fig:exclusion}
\end{figure}

Fig.~\ref{fig:summe} shows the $1\sigma$ and $2\sigma$
contour lines of the right sign fit for different $|\dm{31}|$
and different $\theta_{13}$ in the case with no charge separation 
(i.e. muons of either charge coming from $\nu_\mu$ or $\bar{\nu}_\mu$). 
With decreasing $\theta_{13}$, the precision in the determination 
of $\sin^2 2\theta_{13}$ gets worse, but this method allows to 
determine $\sin^2 2\theta_{13}$ already down to values of order 
$\mathcal{O}(10^{-2})$ (see also fig.~\ref{fig:exclusion} and 
discussion). Since the fitted spectrum includes the dominant
disappearance channel, quite a good precision can be obtained also
for $|\dm{31}|$.
\begin{figure}[htb!]
\begin{center}
\epsfig{file=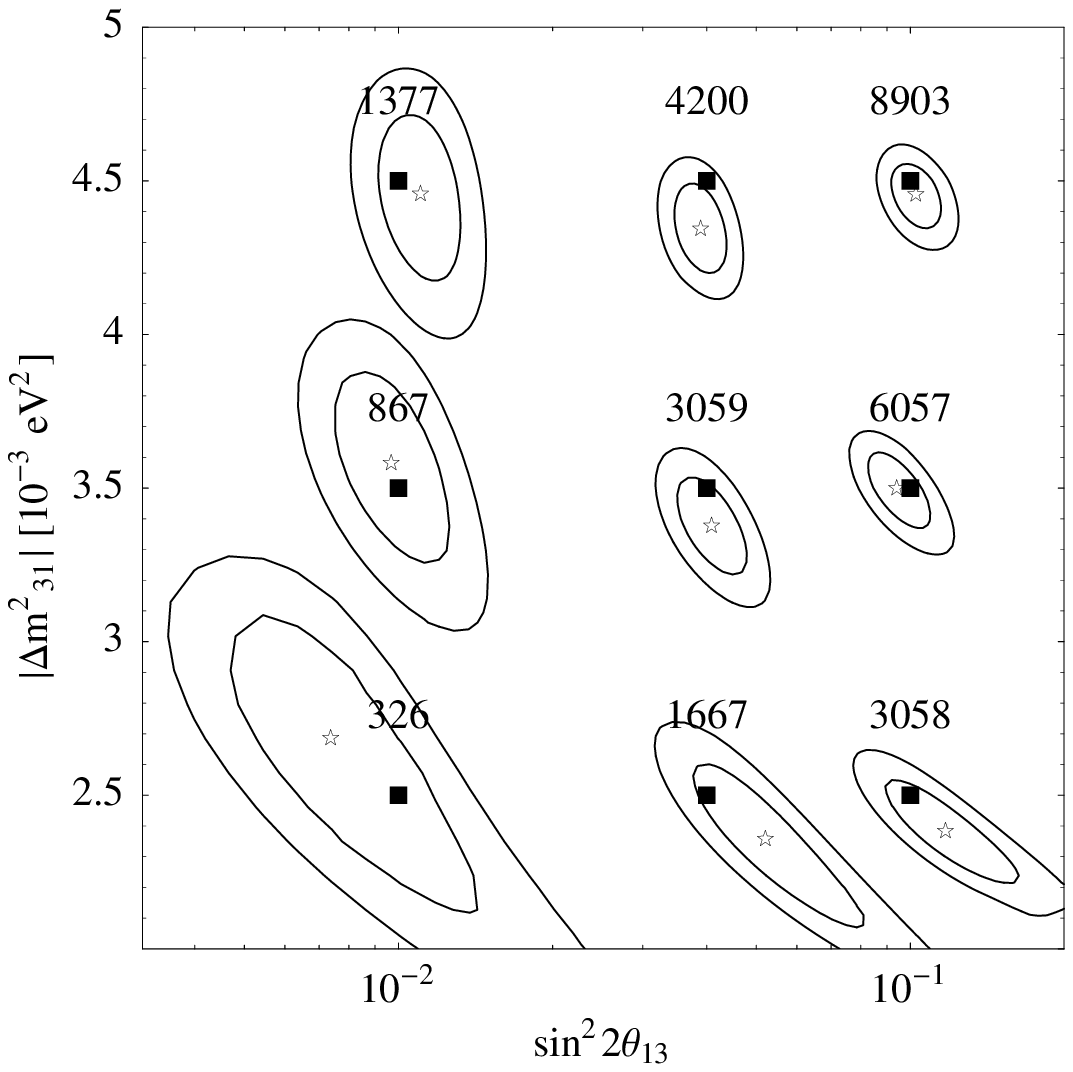,width=10cm}
\end{center}
\mycaption{Fit to the muon neutrino appearance ($\reub$) spectrum 
with charge identification for $\dm{31}>0$ at a baseline of 7332~km. 
Shown are the $1\sigma$ and $2\sigma$ contour lines. The black 
rectangles denote the parameter pair for which the data were generated 
and the stars denote the best fit. The numbers next to the contour 
lines are $\chi^2$ per 2~d.o.f values for the local minima of the same
parameters, but with the wrong sign of $\dm{31}$. The global minimum 
is excluded in this case because its $\dm{31}$ is already excluded.}
\label{fig:appearance}
\end{figure}

Charge separation capabilities improve the sensitivity for
$\sin^2 2\theta_{13}$. This is shown in fig.~\ref{fig:appearance}, 
which shows the results of a fit to the appearance channel only 
(wrong sign muon events). In this case, which does not include 
the disappearance events in the analysis, one looses of course 
precision in finding the right value of $|\dm{31}|$. The precision 
for $|\dm{31}|$ can of course be improved in this case by a 
separate analysis of the disappearance spectrum \cite{BARGER99b}.
Fig.~\ref{fig:appearance} shows that charge identification improves
the precision in the determination of $\sin^2 2\theta_{13}$ for 
small values of $\sin^2 2\theta_{13}$ and the sensitivity will
go down to values of order $\mathcal{O}(10^{-3} - 10^{-4})$. 
Note, however, that for not too small mixing angles 
$\sin^2 2\theta_{13}$ of about one magnitude below the current 
CHOOZ constraint, we find that the sum-rates without charge 
identification provide at least an equally well suited 
possibility to extract the value of $\sin^2 2\theta_{13}$.
The determination of the sign of $\dm{31}$ is always better 
in the appearance channel, but note that the sum-rates give also 
quite good confidence levels for not too small mixing angles $
\theta_{13}$.

\begin{figure}[htb!]
\begin{center}
\epsfig{file=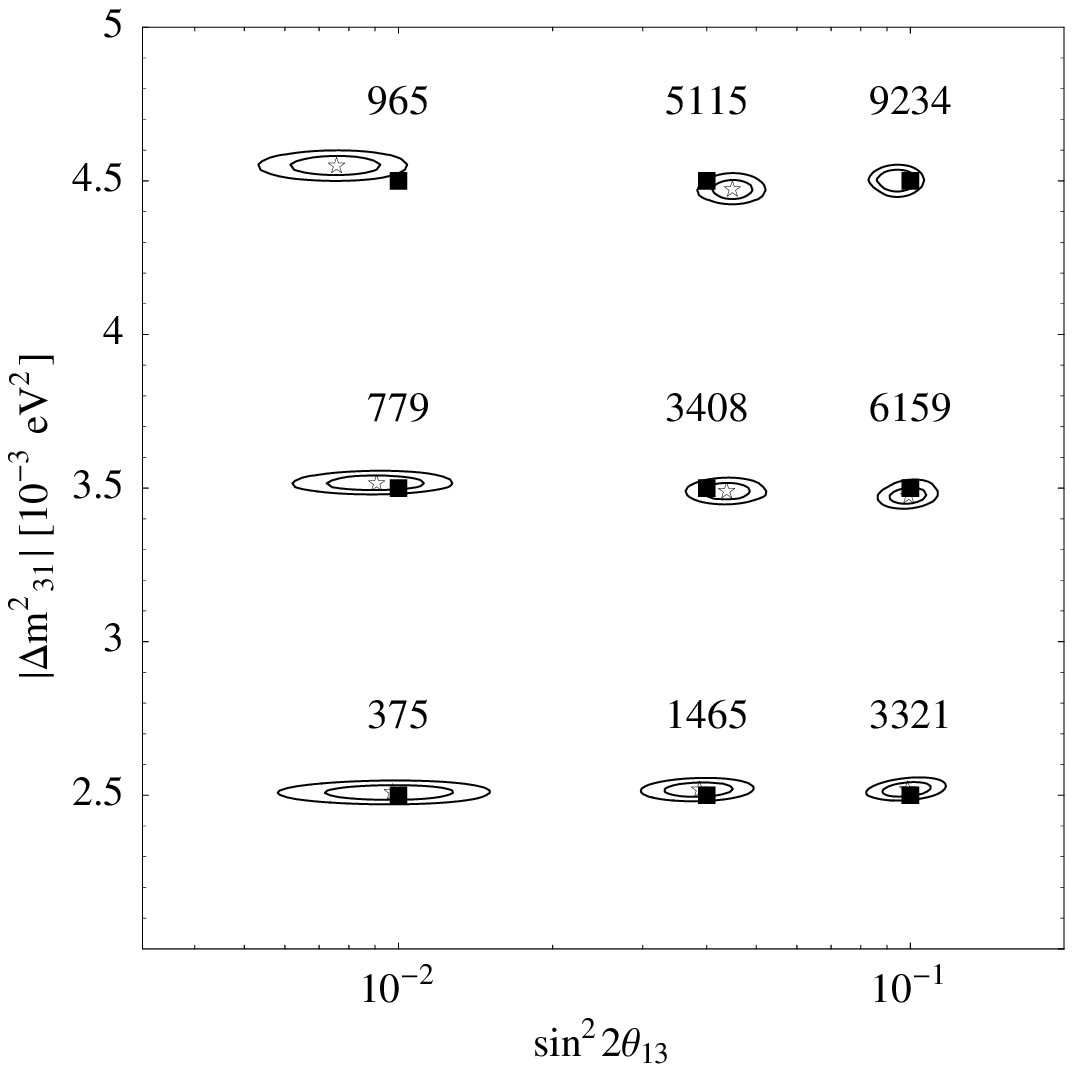,width=10cm}
\end{center}
\mycaption{Fit to the muon neutrino spectrum with charge identification 
(i.e. $\ruu$ and $\reub$ channels separated) for $\dm{31}>0$ at a 
baseline of 7332~km. Shown are the  $1 \sigma$ and
$2 \sigma$ contours. The rectangles denote the parameter pairs 
for which the data were generated and the stars denote the best 
fits. The numbers printed next to the contour lines are the 
local minima of $\chi ^2$ per 2~d.o.f with the same parameters, but
the wrong sign of $\dm{31}$. The global minimum is excluded 
because its $\dm{31}$ value is out of range.}
\label{fig:combined}
\end{figure}
If charge identification is available then there is however an even 
better strategy by combining all available information, i.e. a
global fit to the matter effects of both the appearance and disappearance
channels. The results of this fit are shown in fig.~\ref{fig:combined}. 
Now high precision is obtained both for $\theta_{13}$ and in the 
determination of $\dm{31}$.

The presented fits were all done with a fixed mixing angle 
$\sin^2 2\theta_{23} = 1$ and it is in principle no problem to 
perform a full three parameter fit for $\dm{31}$, 
$\sin^2 2\theta_{13}$ and $\sin^2 2\theta_{23}$. The contour lines
of our fits have to be extended into the third 
$\sin^2 2\theta_{23}$ dimension and the precision is roughly
given by the results in ref.~\cite{BARGER99b}. As explained above
in the analytic discussion, there is no need to include further 
parameters like $\dm{21}$ and the CP-phase $\delta$ as long as 
the LMA~MSW solution is not realized with a value of 
$\dm{21}$ close to its current upper limit of order 
$\mathcal{O}(10^{-4})$ \cite{DFLR,FLPR,romanino,gavela-b,KuoP87,sato1}.
If this scenario were however realized, such that CP-effects play a 
role, then this would be another reason to go for measurements based
on matter effects to largest possible baselines like $7332$~km. The 
point is that the relative magnitude of effects coming from 
$\dm{21}\neq 0$ and $\delta$ decrease for longer baselines \cite{FLPR}. 
Such a scenario would however also enable a measurement of the CP 
violating effects at shorter baselines such that two experiments,
one at a large baseline (for matter effects) and another a shorter 
baseline (say 2800~km for CP violating effects) would be ideal.

Fig. \ref{fig:exclusion} shows finally the region of the 
$\dm{31}$-$\sin^2 2\theta_{13}$ parameter space where a determination
of the sign of $\dm{31}$ will be possible. Shown are the  
$1\sigma$, $2\sigma$ and $3\sigma$ contour lines of a fit with the 
wrong sign of $\dm{31}$ for positive $\dm{31}$ (left plot) and
negative $\dm{31}$ (right plot). 
For positive $\dm{31}$ the cross section favored neutrino channels 
show resonant matter effects whereas for negative $\dm{31}$ the 
cross section disfavored antineutrino channels are resonant. 
Thus in the case of positive $\dm{31}$ the confidence level at 
which the wrong sign can be excluded is slightly better. 
The dotted lines show the limits which can be obtained with
charge identification from the spectral data of the appearance 
channels. The solid lines are based on an analysis of the combined 
channels without charge identification. Charge separation capabilities
which allow to use the statistically favored pure appearance channel 
are thus very important if $\sin^2 2\theta_{13}$ turns out to be 
smaller than $\simeq 10^{-2}$.

From this discussion it seems clear that one should include charge
identification. But there are other issues which may turn out to be
equally important as the discussion of statistics and rates which
was presented here. If, for example, right sign muon rejection 
(i.e. charge identification) turns out to be less good than hoped,
or if there were background issues which make it hard or impossible
to isolate correctly the wrong sign $\reu$ muon signal then our 
proposed method without charge identification should still work. 
Many of these issues are still under discussion 
\cite{FNALworkshopB,FNALworkshopC,FNALworkshopD}. 
To illustrate the requirements for charge identification 
assume 100,000 muon events in a detector (which corresponds roughly
to a baseline of $3000$~km and a beam energy of $50$~GeV with the
usual beam and detector parameters) and right sign charge rejection 
of $10^{-4}$, which would lead in average to 10 background
events in the appearance channel which limits the ultimate sensitivity 
to $\sin^2 2 \theta_{13}$ considerably. 
The problem can not be overcome by increasing statistics (i.e. by 
increasing $N_{\mu^\pm} \NKT \epsilon_\mu$) since the signal to 
background ratio stays constant. Any real detector may thus be 
limited by its right sign charge rejection capability
and the number of right sign muons. This is also connected to the
question which baseline should be used. The amount of background
is typically directly proportional to the disappearance rates. 
This would favour larger baselines, since the disappearance 
rates drop faster than the appearance rates, improving the signal 
to background ratio for longer baselines. The decreased statistics 
in the $\sin^2 2 \theta_{23}$ and $|\dm{31}|$ determination 
(with a precision of the order of a few percent) is not a problem 
since this will be limited by systematics and not statistics, 
even for the longest baselines.


\section{Conclusions \label{sec:concl}}

We studied in this paper different possibilities to use matter
effects to determine the value and the sign of $\dm{31}$ and 
the magnitude of $\sin^2 2\theta_{13}$ in very long baseline 
experiments with neutrino factories. The analysis rests on the 
detection of muons coming from muon-neutrinos or antineutrinos
and we distinguish detectors with and without right 
sign muon charge rejection capabilities. This may be important 
since the $\reu$ and $\reub$ appearance channels have rather 
small total event rates and they require very good efficiencies 
for the detection of wrong sign muon events. Backgrounds due to 
charge misidentification play an especially important role
for very small $\sin^2 2 \theta_{13}$ and at too short baselines, 
where the dominant non-oscillated neutrino rates are high.
Studies of charge identification capabilities are presently 
performed \cite{gavela-d,FNALworkshopB,FNALworkshopC,FNALworkshopD}.

We discussed the relevant appearance and disappearance rates 
analytically and we performed numerical simulations which
were used to test the parameter extraction from event rates
in a statistical reliable way. Our results show for large 
baselines like $7332$~km that there are matter effects of 
comparable size in the event rates in the $\reu$ and $\reub$ 
appearance and in the $\ruu$ and $\ruub$ disappearance channels. 
Matter effects in the disappearance 
channels alone are however statistically somewhat disfavored, 
since they have to be extracted from the relatively large 
amount of un-oscillated events. Without sufficient charge 
identification capabilities there is however still the advantage 
that matter effects in the combined muon rates resulting
from $\nu_\mu$ and $\bar{\nu}_\mu$, i.e. generated from the 
combined appearance and disappearance channels, add in a 
constructive way. This allows a determination of 
$\sin^2 2\theta_{13}$ and the sign of $\dm{31}$ even without 
charge identification down to mixings 
$\sin^2 2\theta_{13} = 10^{-2}$ and with a precision comparable 
to the appearance channels with perfect charge separation.
This new method allows thus to measure or limit the mixing 
$\sin^2 2\theta_{13}$ roughly one order of magnitude below 
the present CHOOZ limit.
With charge identification one can get better results from
the appearance channels alone. We pointed however out that 
with charge identification the best results can be obtained 
by combining all information from the appearance and 
disappearance channels. In this case one can measure or limit 
the mixing down to $\sin^2 2\theta_{13} \simeq 10^{-4}$.
We gave exclusion plots which display the parameter range
where a determination of the sign of the mass squared 
difference $\dm{31}$ will be possible. 

Our results were obtained in the approximation where $\dm{21}$ 
corrections are negligible which allowed an analytic description. 
In the limit $\dm{21} = 0$ CP violating effects drop out and 
matter effects can be understood in an effective two neutrino 
scheme which is inserted into the full three neutrino oscillation
formulae. It should however be stressed that there will be 
sizable corrections to this picture for shorter baselines
if $\dm{21}$ is at its upper limit. In this case the fit of 
$\dm{31}$, $\sin^2 2\theta_{13}$ and $\sin^2 2\theta_{23}$ had
to include in addition $\dm{21}$ and the CP-phase $\delta$.
These extra $\dm{21}\neq 0$ effects become however much smaller
for longer baselines, such that larger baselines are safer 
without loosing detection capabilities for matter effects.
We did not discuss in more detail baseline and beam energy 
optimization since it depends on the preferred quantities, 
improvements on the knowledge of some parameters and the number 
of different baselines (i.e. beams and experiments) available.
We mentioned however that a too short baseline could potentially 
also be dangerous if one neglects imperfect charge separation 
and other backgrounds. It seems therefore that larger baselines
above 3000~km are also preferred from this point of view.
We restricted our analysis therefore to a baseline of 7332~km 
which is very well suited for our study. We used a muon beam 
energy of $20~\GeV$, since it gives a spectrum centered around 
the resonance energy of matter effects in earth and which seems 
to be preferred by recent studies of entry level neutrino 
factories \cite{FNALworkshopA}.


\vspace*{7mm}
Acknowledgments: We wish to thank S.T.~Petcov and A.~Romanino for 
discussions on subjects related to this work.


\newpage

\bibliographystyle{phaip}

\end{document}